# Particle-Number Projected Hartree-Fock-Bogoliubov Study with Effective Shell Model Interactions

I. Maqbool,[1] J.A. Sheikh,[1,2] P.A. Ganai,[1] and P. Ring[3]

[1]*Department of Physics, University of Kashmir, Srinagar 190 006, India*

[2]*Department of Physics and Astronomy, University of Tennessee Knoxville, TN 37996, USA*

[3]*Physik-Department, Technische Universität München, D-85747 Garching, Germany*

We perform particle-number projected mean-field study using the recently developed symmetry-projected Hartree-Fock-Bogoliubov (HFB) equations. Realistic calculations have been performed in sd- and fp-shell nuclei using the shell model empirical intearctions, USD and GXPFIA. It is demonstrated that the mean-field results for energy surfaces, obtained with these shell model interactions, are quite similar to those obtained using the density functional approaches. Further, it is shown that particle-number projected results, for neutron rich isotopes, can lead to different ground-state shapes in comparison to the bare HFB calculations.

## I. INTRODUCTION

One of the primary research goals in nuclear structure physics is to consider correlations going beyond the mean-field approximation [1]. The mean-field theory in the form of Hartree-Fock (HF) and Hartree-Fock-Bogoliubov (HFB) approaches have been quite successful in describing the gross features of atomic nuclei [2–6]. In particular, mean-field study with density dependent effective interactions, which include Skyrme [7–10], Gogny [11–13] and relativistic models [14–16] have provided an accurate description of the ground-state properties of atomic nuclei ( binding energies, deformations, radii and etc.). However, it is well recognized that mean-field approximation breaks down, in particular, when approaching the limits of spin, particle stability and isospin. In approaching these limits, the correlations going beyond the mean-field approximation play a pivotal role. The related problem is that the product wavefunction ansatz employed in the mean-field theory breaks the symmetries that original many-body Hamiltonian obeys [1, 4]. These broken symmetries, for instance, include gauge-symmetry associated with the particle-number, rotational and isospin symmetries.

Recently, it has been demonstrated that the variation of the projected energy functional, which can be completely expressed in terms of the bare HFB densities of $\rho$ and $\kappa$, results into HFB like equations [17]. The only difference is that the expressions for the fields become more involved and depend on the conserved symmetries that are required to be restored. The major advantage of this so-called symmetry-projected HFB (SP-HFB) equations is that the existing programs solving bare HFB equations can be used and only the expressions for the fields need to be redefined. As a matter of fact, this approach has been already implemented in the Skyrme and the relativistic density functional approaches [18–20]. However, in the application of the projected HFB formalism to the density functional theories, one encounters the prob-lem of poles. This problem has been known for quite sometime and has recently been a subject of intensive study in nuclear theory [13, 21–25]. This problem exists because Hartree-Fock and the pairing fields are usually obtained from different interactions. In most of the density functional theories, except in Gogny, Hartree Fock (HF) or particle-hole part of the density functional are obtained in a least-squares fit to the known properties of the closed shell nuclei and nuclear matter. The interaction in the pairing or particle-particle channel is then assumed to be of phenomenological form with the parameters adjusted to the odd-even mass differences. For the case of Gogny force, it has been shown that this problem of pole can be avoided only if all the exchange terms appearing in the density functional are included [13]. Recently, it has been demonstrated, for density functionals with density-dependence of polynomial type, that the problem of pole can be cured by correcting for the self-interaction terms[25]. However, in most of the realistic density functional, the dependence is of fractional power and is mandatory to obtain the correct saturation property.

The alternative to the density functional approach is to perform the symmetry projection with the interaction defined in the shell model configuration space. The interaction can be obtained for a given configuration space following the standard G-matrix renormalization procedure [26–28]. The projection analysis with interaction defined in the shell model configuration space has been performed, quite sometime ago, with a simplified pairing plus quadrupole-quadrupole interaction and considering three major oscillator shells [29–34]. The significant departure of the present study from this earlier work is that we shall consider more realistic effective interactions and also include all the exchange terms. Furthermore, in the present study, we shall employ SP-HFB equations to perform the projection as they are numerically easier to solve and correspond to non-linear diagonalization of a matrix. In the earlier work, the gradient methods were employed which are very difficult to implement numerically [33, 35].

In our earlier study, we have solved SP-HFB equations for particle-number projection analysis by considering a simplified model of a deformed single-j shell. It was shown that the numerical work involved in solving SP-HFB equations is a factor of 2-3 larger than solving the bare HFB equations and further the projected calculations were shown to be almost identical with the exact results. In the present article, we shall perform projected HFB study of sd- and fp-shell nuclei with realistic interactions. It is known that shell model interactions, USD [36, 37], KB3 [38, 39], GXPF1 [40, 41] and GXPF1A [42] are very successful in reproducing the experimental data in these two regions. Although, these interactions have been derived using the interacting shell model technique by fitting the experimental data, and are appropriate in the shell model context. We shall employ USD and GXPF1A interactions in the projected HFB study as the main theme of the projection theory is to include correlations going beyond the mean-field level and these should be a part of the correlations considered in the interacting shell model approach. After all, performing full projection and then mixing these projected states using generator coordinate method should bring the mean-field analysis closer to the interacting shell model results.

The present article is organized in the following form : In the next section, we first present some basic quantities involved in the mean-field method, for completeness, starting with the shell model Hamiltonian. The projected HFB framework is then briefly presented in section III. In our earlier work [17, 43], the projected expressions were derived only for the identical particles. In this section, we shall present the projection formalism for a generalized nuclear Hamiltonian, which includes the neutron-proton interaction term. The results obtained in HFB and PHFB study for sd- and fp-shell will be discussed in section IV and finally the present work is summarized in section V.

## II. BASIC FORMULAE

The nuclear Hamiltonian consisting of neutron-neutron, proton-proton and neutron-proton interaction terms is given by

$$\hat{H} = \hat{H}_n + \hat{H}_p + \hat{H}_{np} \ , \tag{1}$$

where,

$$\hat{H}_t = \hat{T}_t + \hat{V}_t \tag{2}$$

$$\hat{H}_t = \sum_{t_1} \epsilon_{t_1} a_{t_1}^\dagger a_{t_1} - \sum_{\substack{t_1 t_2 t_3 t_4 J_t \\ t_1 \le t_2, t_3 \le t_4}} (2J_t+1)^{1/2} < t_1 t_2 |\hat{v}| t_3 t_4 >_{J_t} \frac{1}{\sqrt{(1+\delta_{t_1 t_2})(1+\delta_{t_3 t_4})}} \left( (a_{t_1}^\dagger a_{t_2}^\dagger)_{J_t} \times (\tilde{a}_{t_3} \tilde{a}_{t_4})_{J_t} \right)_{00} \tag{3}$$

where "t" is equal to n(p) for neutrons (protons) and the labels "$t_1, t_2, ...$" denote the angular-momentum and other quantum numbers necessary to define a spherical single-particle state uniquely. $a^\dagger(a)$ are the nucleon creation (annihilation) operators and $\tilde{a}_{j_t m_t} = (-1)^{j_t + m_t} a_{j_t - m_t}$. $< t_1 t_2 |\hat{v}| t_3 t_4 >_{J_t}$ denotes the anti-symmetric normalized two-body matrix element. For performing the projected mean-field analysis, it is required to first express the two-body term in Eq. (3) in the following standard un-coupled representation

$$\hat{V}_t = \sum_{\substack{t_1 t_2 t_3 t_4 \\ m_{t_1} m_{t_2} m_{t_3} m_{t_4}}} < t_1 m_{t_1}, t_2 m_{t_2} |\hat{v}| t_3 m_{t_3}, t_4 m_{t_4} > a_{t_1 m_{t_1}}^\dagger a_{t_2 m_{t_2}}^\dagger a_{t_3 m_{t_3}} a_{t_4 m_{t_4}} \ , \tag{4}$$

where the un-coupled two-body matrix element is given by

$$< t_1 m_{t_1}, t_2 m_{t_2} |\hat{v}| t_3 m_{t_3} t_4 m_{t_4} > = - \sum_{J_t M_t} < t_1 t_2 |\hat{v}| t_3 t_4 >_{J_t} \frac{\sqrt{(1+\delta_{t_1 t_2})(1+\delta_{t_3 t_4})}}{4}$$
$$\begin{bmatrix} j_{t_1} & j_{t_2} & J_t \\ m_{t_1} & m_{t_2} & M_t \end{bmatrix} \begin{bmatrix} j_{t_3} & j_{t_4} & J_t \\ m_{t_3} & m_{t_4} & M_t \end{bmatrix} \ . \tag{5}$$

The symbol [...] in the above equation denotes the Clebsch-Gordon coefficient. The neutron-proton interaction term can be similarly expressed in the un-coupled representation as

$$\hat{H}_{np} = 4 \sum_{\substack{n_1 n_2 p_1 p_2 \\ m_{n_1} m_{n_2} m_{p_1} m_{p_2}}} < n_1 m_{n_1}, p_1 m_{p_1} |\hat{v}| n_2 m_{n_2}, p_2 m_{p_2} > a_{n_1 m_{n_1}}^\dagger a_{p_1 m_{p_1}}^\dagger a_{n_2 m_{n_2}} a_{p_2 m_{p_2}} \ , \tag{6}$$

where $n_1, n_2$ $(p_1, p_2)$ denote the quantum numbers for neutrons (protons).

The most basic of mean-field approaches to describe the pairing-correlations of the Hamiltonian (1) is the HFB method. In this method the ground-state wave-function is a product state of quasiparticle operators $(\alpha, \alpha^\dagger) = (\alpha_1, \ldots, \alpha_M; \alpha_1^\dagger, \ldots, \alpha_M^\dagger)$. These quasiparticle operators are connected to the original particle operators by a linear transformation. The quasiparticle transformation is defined independently for neutrons and protons and is given by

$$\alpha_k = \sum_{t_1} \left( U_{t_1 k}^* a_{t_1} + V_{t_1 k}^* a_{t_1}^\dagger \right), \tag{7}$$

$$\alpha_k^\dagger = \sum_{t_1} \left( V_{t_1 k} a_{t_1} + U_{t_1 k} a_{t_1}^\dagger \right). \tag{8}$$

These equations can be rewritten in the matrix form as

$$\begin{pmatrix} \alpha \\ \alpha^\dagger \end{pmatrix} = \begin{pmatrix} U^\dagger & V^\dagger \\ V^T & U^T \end{pmatrix} \begin{pmatrix} a \\ a^\dagger \end{pmatrix} = \mathcal{W}^\dagger \begin{pmatrix} a \\ a^\dagger \end{pmatrix}. \tag{9}$$

The quasiparticle operators need to satisfy the same fermion commutation relations as the original operators and, therefore, the transformation matrix is required to be unitary

$$\mathcal{W}^\dagger \mathcal{W} = \mathcal{W}\mathcal{W}^\dagger = I, \tag{10}$$

which leads to following relations among the coefficients $U$'s and $V$'s

$$U^\dagger U + V^\dagger V = I, \qquad UU^\dagger + V^* V^T = I, \tag{11}$$
$$U^T V + V^T U = 0, \qquad UV^\dagger + V^* U^T = 0. \tag{12}$$

The quasiparticle operators annihilate the quasiparticle vacuum $|\Phi\rangle$, defined by

$$\alpha_k |\Phi\rangle = 0, \tag{13}$$

for all $k$. In mean-field theory $|\Phi\rangle$ represents an approximation to the ground-state of the system and turns out to be a generalized Slater-determinant [1]. Since the quasiparticle transformation mixes creation and annihilation operators, $|\Phi\rangle$ does not correspond to a wavefunction with good particle-number. Therefore, we have two types of densities, the normal density, $\rho$ and the pairing-tensor, $\kappa$, which are defined as

$$\rho_{t_1 t_2} = \langle \Phi | a_{t_2}^\dagger a_{t_1} | \Phi \rangle, \qquad \kappa_{t_1 t_2} = \langle \Phi | a_{t_2} a_{t_1} | \Phi \rangle. \tag{14}$$

These can be expressed in terms of the HFB coefficients as

$$\rho = V^* V^T, \qquad \kappa = V^* U^T = -UV^\dagger. \tag{15}$$

The HFB energy is given by

$$E_{HFB} = \frac{\langle \Phi | \hat{H} | \Phi \rangle}{\langle \Phi | \Phi \rangle}, \tag{16}$$

$$= \sum_t \left[ \text{Tr}((\epsilon^t + \frac{1}{2}\Gamma^t)\rho^t - \frac{1}{2}(\Delta^t \kappa^{t*})) \right]$$
$$+ \frac{1}{2}\text{Tr}(\Gamma^p \rho^n + \Gamma^n \rho^p), \tag{17}$$

where the HFB fields are defined as

$$\Gamma_{t_1 t_3}^t = \sum_{t_2 t_4} \overline{v}_{t_1 t_2 t_3 t_4} \rho_{t_4 t_2}^t, \tag{18}$$

$$\Delta_{t_1 t_2}^t = \frac{1}{2} \sum_{t_3 t_4} \overline{v}_{t_1 t_2 t_3 t_4} \kappa_{t_3 t_4}^t. \tag{19}$$

The variation of the HFB energy functional Eq. (16) results in the standard HFB equations for neutrons and protons separately

$$\mathcal{H}^t \begin{pmatrix} U^t \\ V^t \end{pmatrix} = E_i^t \begin{pmatrix} U^t \\ V^t \end{pmatrix}, \tag{20}$$

where

$$\mathcal{H}^t = \begin{pmatrix} \epsilon^t + \Gamma^t - \lambda^t & \Delta^t \\ -\Delta^{t*} & -(\epsilon^t + \Gamma^t)^* + \lambda^t \end{pmatrix}. \tag{21}$$

The HFB wavefunction does not have a well defined particle number and the Lagrangian parameter ($\lambda^t$) is introduced to have the correct particle-number on the average. The diagonalisation of the HFB matrix (21) gives rise to the quasiparticle energies ($E_i^t$) and the excitation spectrum of the many-body system is constructed from these energies.

## III. PARTICLE-NUMBER PROJECTION FORMALISM

The projected-energy functional with well defined neutron (N) and proton (Z) number is given by

$$E^P = \frac{\left\langle \Phi | \hat{H} \hat{P}^N \hat{P}^Z | \Phi \right\rangle}{\left\langle \Phi | \hat{P}^N \hat{P}^Z | \Phi \right\rangle} = \frac{\int d\phi_n \int d\phi_p \langle \Phi | \hat{H} e^{i\phi_n(\hat{N}-N)} e^{i\phi_p(\hat{Z}-Z)} | \Phi \rangle}{\int d\phi_n \int d\phi_p \langle \Phi | e^{i\phi_n(\hat{N}-N)} e^{i\phi_p(\hat{Z}-Z)} | \Phi \rangle}, \tag{22}$$

where the particle-number projection operator is defined as

$$P^T = \frac{1}{2\pi} \int d\phi \, e^{i\phi_t(\hat{T}-T)}. \tag{23}$$

where $T = N(Z)$ denotes the neutron (proton) particle-number. It has been demonstrated in ref. [17] that the variation of the projected-energy results in equations that have same structure as that of standard HFB like equations. The projected HFB equation is given by [17]

$$\mathcal{H}^T \begin{pmatrix} U^t \\ V^t \end{pmatrix} = \mathcal{E}_i^t \begin{pmatrix} U^t \\ V^t \end{pmatrix},\tag{24}$$

where

$$\mathcal{H}^T = \begin{pmatrix} \varepsilon^T + \Gamma^T + \Lambda^T - \lambda^t & \Delta^T \\ -(\Delta^T)^* & -(\varepsilon^T)^* - (\Gamma^T)^* - (\Lambda^T)^* + \lambda^t \end{pmatrix}.\tag{25}$$

The number-projected expressions for the fields are given by

$$\begin{aligned}
\varepsilon^T &= \frac{1}{2}\int d\phi_t\ y^t(\phi_t)\left\{Y^t(\phi_t)\mathrm{Tr}[\epsilon^t\rho^t(\phi_t)] + [1 - 2ie^{-i\phi_t}\sin\phi_t\rho^t(\phi_t)]\epsilon^t C^t(\phi_t)\right\}\\
&\quad + h.c.
\end{aligned}\tag{26}$$

$$\begin{aligned}
\Gamma^T &= \frac{1}{2}\int d\phi_t\ y^t(\phi_t)\left(Y^t(\phi_t)\frac{1}{2}\mathrm{Tr}[\Gamma^t(\phi_t)\rho^t(\phi_t)] + \frac{1}{2}[1 - 2ie^{-i\phi_t}\sin\phi_t\rho^t(\phi_t)]\Gamma^t(\phi_t)C^t(\phi_t)\right)\\
&\quad + \frac{1}{2}\int d\phi_t\int d\phi_{t'}y^t(\phi_t)y^{t'}(\phi_{t'})\left(Y^t(\phi_t)\mathrm{Tr}[\Gamma^{tt'}(\phi_{t'})\rho^t(\phi_t)] + [1 - 2ie^{-i\phi_t}\sin\phi_t\rho^t(\phi_t)]\Gamma^{tt'}(\phi_{t'})C^t(\phi_t)\right)\\
&\quad + h.c.
\end{aligned}\tag{27}$$

$$\begin{aligned}
\Lambda^T &= -\frac{1}{2}\int d\phi_t\ y^t(\phi_t)\left(Y^t(\phi_t)\frac{1}{2}\mathrm{Tr}[\Delta^t(\phi_t)\overline{\kappa^t}^*(\phi_t)] - 2ie^{-i\phi_t}\sin\phi_t\ C(\phi_t)\Delta(\phi_t)\overline{\kappa_t}^*\right)\\
&\quad + h.c.
\end{aligned}\tag{28}$$

$$\Delta^T = \frac{1}{2}\int d\phi_t\ y^t(\phi_t)e^{-2i\phi_t}C^t(\phi_t)\Delta^t(\phi_t) - (..)^{\mathsf{T}}.\tag{29}$$

In Eq. (27), $t'$ is equal to $n(p)$ for $t$ equal to $p(n)$. Further, the rotated fields in the above equations are given by

$$\Gamma_{t_1 t_3}^t(\phi_t) = \sum_{t_2 t_4}\overline{v}_{t_1 t_2 t_3 t_4}\rho_{t_4 t_2}^t(\phi_t),\tag{30}$$

$$\Gamma_{t_1' t_3'}^{tt'}(\phi_t) = \sum_{t_2 t_4}\overline{v}_{t_1' t_2 t_3' t_4}\rho_{t_4 t_2}^t(\phi_t),\tag{31}$$

$$\Delta_{t_1 t_2}^t(\phi_t) = \frac{1}{2}\sum_{t_3 t_4}\overline{v}_{t_1 t_2 t_3 t_4}\kappa_{t_3 t_4}^t(\phi_t),\tag{32}$$

$$(\overline{\Delta}_{t_3 t_4}^t(\phi_t))^* = \frac{1}{2}\sum_{t_1 t_2}\overline{\kappa}_{t_1 t_2}^{t*}(\phi_t)\overline{v}_{t_1 t_2 t_3 t_4},\tag{33}$$

$$\rho^t(\phi_t) = C^t(\phi_t)\rho^t,\tag{34}$$

$$\kappa^t(\phi_t) = C^t(\phi_t)\kappa^t = \kappa^t(C^t(\phi_t))^{\mathsf{T}},\tag{35}$$

$$\overline{\kappa}^t(\phi_t) = e^{2i\phi_t}\kappa^t(C^t(\phi_t))^* = e^{2i\phi_t}(C^t(\phi_t))^\dagger\kappa^t,\tag{36}$$

$$C^t(\phi_t) = e^{2i\phi_t}\left(1 + \rho^t(e^{2i\phi_t} - 1)\right)^{-1},\tag{37}$$

$$x^t(\phi_t) = \frac{1}{2\pi}\frac{e^{i\phi_t(T)}\det(e^{i\phi_t})}{\sqrt{\det C^t(\phi_t)}},\tag{38}$$

$$y^t(\phi_t) = \frac{x^t(\phi_t)}{\int d\phi_t\ x^t(\phi_t)},\quad \int d\phi_t\ y(\phi_t) = 1,\tag{39}$$

and

$$Y^t(\phi_t) = ie^{-i\phi_t}\sin\phi_t\ C^t(\phi_t) - i\int d\phi_t'y^t(\phi_t')e^{-i\phi_t'}\sin\phi_t'\ C^t(\phi_t').\tag{40}$$

It needs to be stressed that as compared to the HFB case, the projected quasiparticle energies obtained from the diagonalisation of the projected HFB matrix Eq. (25) have no meaning. The only meaningful quantity is the total projected-energy which is given by

$$E_{tot} = \int d\phi_n \int d\phi_p \; y^n(\phi_n) y^p(\phi_p) \hat{H}(\phi_n \phi_p), \tag{41}$$

$$= \sum_t \left( \int d\phi_t \; y^t(\phi_t) \left\{ H_{sp}^t(\phi_t) + H_{ph}^t(\phi_t) + H_{pp}^t(\phi_t) \right\} \right)$$

$$+ \int d\phi_n \int d\phi_p \; y^n(\phi_n) y^p(\phi_p) H_{ph}^{np}(\phi_n \phi_p), \tag{42}$$

where,

$$H_{sp}^t(\phi_t) = \mathrm{Tr}\left[ \epsilon^t \rho^t(\phi_t) \right], \tag{43}$$

$$H_{ph}^t(\phi_t) = \frac{1}{2} \mathrm{Tr}\left[ \Gamma^t(\phi_t) \rho^t(\phi_t) \right], \tag{44}$$

$$H_{ph}^{np}(\phi_n \phi_p) = \frac{1}{2} \mathrm{Tr}\left[ \Gamma^{np}(\phi_p) \rho^n(\phi_n) + \Gamma^{pn}(\phi_n) \rho^p(\phi_p) \right], \tag{45}$$

$$H_{pp}^t(\phi_t) = -\frac{1}{2} \mathrm{Tr}\left[ \Delta^t(\phi_t) (\overline{\kappa}^t(\phi_t))^* \right]. \tag{46}$$

In the present work, we will discuss the pairing-energy ($E_{pair}$) and the intrinsic quadrupole momentum ($Q_0$) quite extensively. The expressions for these quantities are given by

$$E_{pair} = \sum_t \left( \int d\phi_t H_{pp}^t(\phi_t) \right), \tag{47}$$

$$Q_0 = \sum_t \left( \int d\phi_t \mathrm{Tr}[\hat{Q}^t \rho^t(\phi_t)] \right), \tag{48}$$

where,

$$\hat{Q}^t = \sqrt{\frac{16\pi}{5}} \; e_t r^2 \hat{Y}_{20}. \tag{49}$$

As is clear from Eq. (25), the projected equations have exactly the same structure as that of normal HFB. Therefore, one can use all the existing HFB computer codes and only the expressions for the projected fields need to be evaluated. The projected fields now involve the integration over the gauge-angle, $\phi$. This integration has been performed using the Gauss-Chebyshev quadrature method. In this method, the integration over the gauge-angle is replaced by a summation. It can be shown [44] that the optimal number of mesh-points in the summation which eliminates all the components having undesired particle numbers is given by

$$M = \max\left( \frac{1}{2} N, \Omega - \frac{1}{2} \right) + 1. \tag{50}$$

where N is the number of particles and $\Omega$ is the degeneracy of the model space.

## IV. RESULTS AND DISCUSSION

In the present work, we have performed HFB and projected HFB study for sd- and fp-shell nuclei using USD [36, 37] and GXPF1A [42] effective interactions, respectively. It is known that spherical shell model analysis performed with these effective interactions provide an accurate description of experimental data. In the mean-field study, we have assumed reflection and axial symmetries as most of nuclei in the two regions are known to obey these symmetries. Furthermore, we shall compare the energy surfaces obtained in the present work with the available results using density functional approaches of Gogny and Skyrme where axial and reflection symmetries have also been enforced.

The calculated total energies ($E^P$) in HFB and PHFB approaches as a function of intrinsic quadrupole moment

$(Q_0)$ are presented in Figs. 1 and 2 for the studied sd-shell region. The intrinsic total quardupole moment is calculated with the effective charges of $e_n = 0.5e, e_p = 1.5e$ and these effective charges have been adopted from the interacting shell model studies [36, 37, 41]. It needs to be stressed here that using these effective charges in the mean-field studies is a questionable issue. In the mean-field study with density functionals, bare charges are employed since the calculations are performed with no core. In the present mean-field study with shell model effective interactions which assume a core, effective charges are obviously required. These effective charges may be different from those used in the shell model analysis as the correlations included in the two approaches are different. We hope to have a better understanding of this problem by performing a systematic analysis for a large class of nuclei in the near future.

Fig. 1 depicts the results of the calculations in both HFB and PHFB approaches for the N=Z sd-shell nuclei, $^{20}$Ne, $^{24}$Mg, $^{28}$Si, $^{32}$S and $^{36}$Ar. The total energy surface for $^{20}$Ne depicts two minima - one for the prolate deformation of $Q_0 = +47.55$ e fm$^2$ and the other for the oblate deformation of $Q_0 = -23.36$ e fm$^2$. The prolate minimum energy is lower than the oblate minimum by about 4 MeV and, therefore, the ground-state shape of $^{20}$Ne is predicted to be prolate. The prolate shape for this system is corroborated by other theoretical approaches and also by the experimental data. The HFB energy surface with the Gogny D1S effective interaction [45] depicts a minimum at the prolate deformation of $Q_0 = +40.15$ e fm$^2$ and an oblate minimum at $Q_0 = -16.06$ e fm$^2$. The prolate minimum is lower than the oblate by about 2 MeV. The calculations using the microscopic-macroscopic approach [46] leads to the minimum with $Q_0 = +25.69$ e fm$^2$. The interacting shell model results with USD interaction and the effective charges of $e_n = 0.35e, e_p = 1.35e$ provide a value of $+52.3$ e fm$^2$ for the intrinsic quadrupole moment of the $2^+$ state with the prescription of ref. [47]. The present results with this set of effective charges gives $Q_0 = +40.42$ e fm$^2$. The corresponding experimental value, mentioned in ref. [47], is $80.7 \pm 10.5$ e fm$^2$.

The HFB and PHFB results for $^{24}$Mg in Fig. 1 again display two minima similar to that of $^{20}$Ne. The minima are at larger deformations of $Q_0 = +58.29$ e fm$^2$ and $-41.11$ e fm$^2$ and further the energy difference between the two minima is about 6 MeV. The HFB study with Gogny interaction predict the two minima at $Q_0 = +56.59$ e fm$^2$ and $-32.64$ e fm$^2$ with the barrier between the two to be approximately 4 MeV high and the microscopic-macroscopic calculations, on the other hand, show the lowest minimum at $Q_0 = +38.09$ e fm$^2$. The extracted value from the shell model calculations is $Q_0 = +57.2$ e fm$^2$ with $e_n = 0.35e, e_p = 1.35e$ and the experimental value is $Q_0 = 63.2 \pm 7.0$ e fm$^2$. The present mean-field calculation with this set of effective charges gives $Q_0 = +49.54$ e fm$^2$. The results for $^{28}$Si, shown in Fig. 1, demonstrate an oblate minimum with $Q_0 =$ $-65.51$ e fm$^2$. The energy surface on the prolate side depicts a shoulder and a minimum in PHFB results depict a very small depth. The Gogny results show a minimum at $Q_0 = -56.28$ e fm$^2$, whereas the microscopic-macroscopic approach predicts the minimum at $Q_0 = -74.57$ e fm$^2$. The shell model and the experimental values for $Q_0$ are $-59.9$ e fm$^2$ and $-56.0 \pm 10.5$ e fm$^2$, respectively. The HFB study with effective charges, $e_n = 0.35e, e_p = 1.35e$, provide $Q_0 = -55.68$ e fm$^2$.

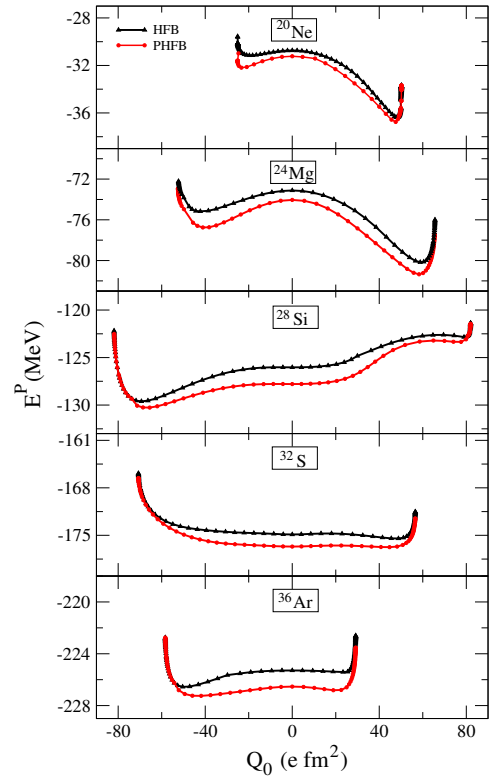

FIG. 1: (Color online) Hartree-Fock-Bogliubov (HFB) and projected HFB (PHFB) results of energy surfaces for the N=Z even-even isotopes in the sd-shell region using USD effective interaction.

The results in Fig. 1 predict spherical shape for $^{32}$S and is consistent with the results obtained with both Gogny and the microscopic-macroscopic approaches. For $^{36}$Ar, the present results show prolate and oblate minima at $Q_0 = +18.87$ e fm$^2$ and $Q_0 = -41.87$ e fm$^2$ with very small barrier between the two. The Hartree Fock + BCS results performed in ref. [48] with the USD interaction also depict a similar energy surface for $^{36}$Ar. The results with Gogny and microscopic-macroscopic models indicate a spherical shape for the lowest configuration.

We have also performed the mean-field study for neutron-rich Ne-isotopes ranging from $^{22}$Ne to $^{28}$Ne. These nuclei have recently attracted considerable attention due to the experimental measurement of the dipole response. Quasiparticle RPA analysis of the dipole response has been performed using relativistic mean-field and Gogny density functionals and it has been shown that the two calculations differ. The reason for the discrepancy can be traced to the two different ground-state shapes obtained in the two approaches. In the relativistic mean-field approach, deformed ground-state solution has been obtained with the pairing-force fitted to the odd-even mass difference. Gogny density functional, on the other hand, leads to the spherical shape with the standard pairing force. Since the ground-state energy minima appears to be sensitive to the pairing interaction for these neutron-rich isotopes, it is of considerable interest to investigate these nuclei in the particle number projected framework. PHFB results, in principle, contain pairing correlations more accurately than the bare HFB calculations and it is ,therefore, expected that HFB and PHFB results should differ for systems for which pairing plays a crucial role.

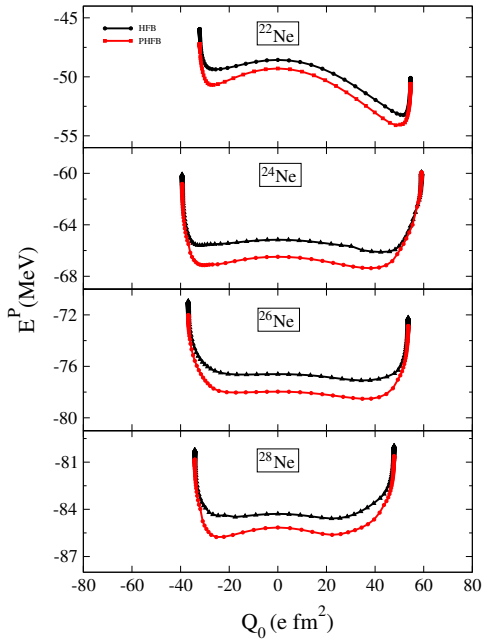

FIG. 2: (Color online) Hartree-Fock-Bogliubov (HFB) and projected HFB (PHFB) results of total energy for neutron rich Ne-isotopes.

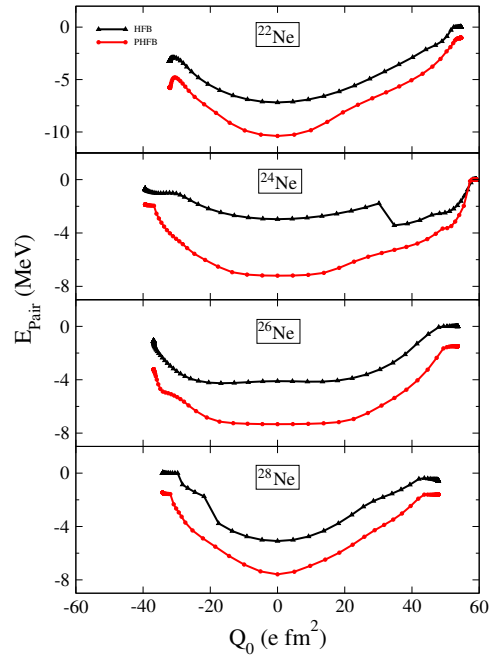

FIG. 3: (Color online) Hartree-Fock-Bogliubov (HFB) and projected HFB (PHFB) results of pairing energy for neutron rich Ne-isotopes.

TABLE I: The average pairing gaps deduced from HFB and PHFB calculations are compared with the empirically obtained pairing gaps from the measured masses.

| Isotope | HFB | | PHFB | | Expt.[49] | |
|---------|------------|------------|------------|------------|------------|------------|
| | $\Delta_N$ | $\Delta_P$ | $\Delta_N$ | $\Delta_P$ | $\Delta_N$ | $\Delta_P$ |
| $^{22}Ne$ | 0.5118 | 0.0000 | 1.4422 | 0.9456 | 2.1975 | 2.6550 |
| $^{24}Ne$ | 1.4087 | 1.1740 | 1.6272 | 1.5180 | 2.0800 | 2.3475 |
| $^{26}Ne$ | 1.0202 | 1.5361 | 1.4577 | 1.9423 | 1.4310 | 2.1435 |
| $^{28}Ne$ | 1.0012 | 1.4791 | 1.1930 | 2.1204 | 1.5050 | 2.1150 |

In Fig. 2, results of total energy surfaces using HFB and PHFB approaches are displayed for the neutron-rich Ne-isotopes. $^{22}$Ne depicts both prolate and oblate minima at $Q_0 = 50$ efm$^2$ and -30 efm$^2$ in both HFB and PHFB approaches. Prolate minimum is lower than the oblate one by about 3 MeV. Shape-coexistence between prolate and oblate shapes is predicted for $^{24}$Ne. In $^{26}$Ne, HFB results show considerable $\gamma$ softness, whereas PHFB results favour a prolate shape for this nucleus. HFB and PHFB results again differ for $^{28}$Ne with HFB predicting shape co-existence and the PHFB results show oblate minimum to be slightly lower than the prolate one. The reason for discrepancy in the HFB and PHFB results is mainly due to the pairing content in the two mean-field solutions. This is demonstrated in Fig. 3, which displays the pairing energy for the studied Ne-isotopes in HFB and PHFB frameworks. The pairing energy in PHFB approach is clearly larger than the corresponding HFB value. In Table I, the average pairing gaps obtained from HFB and PHFB calculations are compared with the corresponding empirical values, obtained from the measured masses. It is quite evident from the comparison that PHFB results are in better agreement with the empirical values.

In the present study, we have also performed the HFB and PHFB calculations for the Cr-isotopes from A=44 to 52 with the GXFP1A effective interaction. We have chosen this isotopic chain as some of these nuclei have been well studied both experimentally and theoretically [50]. Further, this isotopic chain shows a variety of shapes with changing neutron number. The results of the total energy as a function of quadrupole moment are plotted in Fig. 4. $^{44}$Cr is predicted to be spherical in the present analysis and $^{46}$Cr has a prolate minimum at $Q_0 = +79.42$ efm$^2$. It is evident from Fig. 4 that $^{48}$Cr has a well developed deformed minimum at $Q_0 = +119.87$ efm$^2$ and $^{50}$Cr has a minimum at $Q_0 = +111.27$ fm$^2$. $^{52}$Cr in the present analysis is again predicted to be spherical. These results for Cr-isotopes are similar to those obtained with the Gogny force and, as a matter of fact, the shapes of the energy surfaces in Fig. 4 are quite similar to those with the Gogny force [45]. The minima obtained for $^{46}$Cr, $^{48}$Cr and $^{50}$Cr with the Gogny force are approximately at deformation values of $\beta = 0.25$, $0.33$ and $0.26$ and in the present study the corresponding minima are at $\beta = 0.26$, $0.34$ and $0.30$.

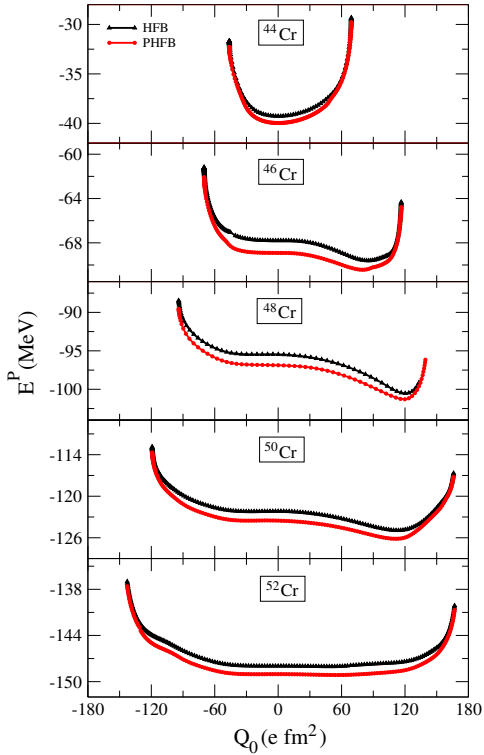

FIG. 4: (Color online) Hartree-Fock-Bogliubov (HFB) and projected HFB (PHFB) results of energy surfaces for even-even Cr-isotopes in using GXFP1A effective interaction.

At first sight, the similarity of the present results with those of Gogny force , in particular for the Cr-isotopes, are quite surprising as the effective forces and the configuration spaces in the two approaches are quite different. In the present study, we have employed the fp-shell space and in the analysis with Gogny force the basis is chosen in such a way that the number of states is 8 times the maximum number of occupied states. However, it is to be noted that GXFP1A effective interaction has been obtained in a least-squares fit to the experimental data in the fp-shell and, therefore, indirectly incorporates the effects of the neglected configuration space. This result is quite encouraging as it implies that the results of the density dependent effective interactions in a larger configuration space can be reproduced with an empirical interaction employing a limited configuration space. Although, performing a mean-field study with a larger configuration space is certainly feasible, but it would cer-tainly be advantageous to have a smaller configuration space when performing symmetry restoration for several quantum numbers.

## V. SUMMARY

The main goal of the present investigation is to include correlations going beyond the standard HFB mean-field approximation. There are primarily two methods of including these fluctuations - one is following the random phase approximation (RPA) path[51, 52] and the other is to perform the projection of the symmetries spontaneously broken by the mean-field approximation [17, 53–55]. Although, both these approaches have been known for quite sometime, but the numerical analysis has been performed only in limited cases. For instance, the RPA study has been performed for spherical nuclei and only recently this method has been applied to deformed systems with axial symmetry [56–58].

The projection of particle number and angular-momentum have been also known for quite sometime and were mostly applied to simplified interactions. The projection method has been applied to Gogny interaction using gradient methods [59, 60], which are very difficult to implement numerically. It has been demonstrated recently that an arbitrary energy functional which is completely expressible in terms of the bare HFB densities results into HFB like equations. It has been shown that projected HFB energy can also be completely expressed in terms of bare HFB densities and, therefore, HFB like equations are also obtained with projected energy functional [17]. The main advantage of this approach, as mentioned earlier, is that one can use the existing HFB codes for performing the projection.

In our earlier work, the projected HFB equations were solved for a simpler model of a deformed single-j shell [17, 43]. In the present work, we have solved these equations for realistic interactions and have performed detailed numerical calculations for sd- and fp-shell nuclei with the empirical shell model interactions. The reason for considering empirical shell model interactions rather than the density dependent effective interactions of Skyrme, Gogny or relativistic mean-field, which are more suited for mean-field study, is that the application of the projection method to density dependent interaction leads to the problem of pole.

It has been shown that deriving Hartree Fock and pairing fields from the same effective interaction and including all the exchange terms, the pole term cancels out. In the present study, we have calculated the mean-fields from the same empirical interactions and also included all the exchange terms and as is evident from the results that pole problem is absent as there are no unphysical kinks in the calculated energy.

It is quite interesting to note from the present work that the results obtained for the energy surfaces, in particular, for the Cr-isotopes are quite similar to those ob-tained with the Gogny interaction and lends support to the notion that density dependent interaction can be replaced by an effective interaction defined over a limited configuration space. Further, the results for energy surfaces are somewhat different in HFB and PHFB studies for the neutron-rich isotopes. The calculated pairing energy in PHFB case is, as expected, larger than the HFB case and drops slowly with increasing deformation. The pairing content in the mean-field solution plays an important role in the study of, for instance, high-spin states.

The present analysis has been limited to lighter nuclei with smaller configuration spaces. In future, we would like to investigate heavier nuclei as mean-field becomes more accurate for heavier systems. The effective interactions for heavier nuclei need to be obtained using G-matrix renormalization procedure. The other possibility is to employ the recently developed mapping procedure [48, 61]. In this method, the parameters of a simplified interaction defined in the shell model configuration space are fitted with the prescription that the energy surface obtained with this interaction is similar to that obtained with density dependent effective interaction. We are presently working along these lines and the results obtained will be discussed in a forthcoming article.